\documentclass[12pt]{article}
\usepackage{epsfig}
\usepackage{graphicx}
\usepackage{amsmath}

\begin{document}

 \newcommand{\beq}{\begin{equation}}
\newcommand{\eeq}{\end{equation}}
\newcommand{\bea}{\begin{eqnarray}} 
\newcommand{\eea}{\end{eqnarray}}
\newcommand{\beqn}{\begin{eqnarray}}
\newcommand{\eeqn}{\end{eqnarray}}
\newcommand{\beas}{\begin{eqnarray*}}
\newcommand{\eeas}{\end{eqnarray*}}
\newcommand{\defi}{\stackrel{\rm def}{=}}
\newcommand{\non}{\nonumber}
\newcommand{\bquo}{\begin{quote}}
\newcommand{\enqu}{\end{quote}}
\newcommand{\qt}{\tilde q}
\newcommand{\m}{\tilde m}
\newcommand{\trho}{\tilde{\rho}}
\newcommand{\tn}{\tilde{n}}
\newcommand{\tN}{\tilde N}
\newcommand{\gsim}{\lower.7ex\hbox{$\;\stackrel{\textstyle>}{\sim}\;$}}
\newcommand{\lsim}{\lower.7ex\hbox{$\;\stackrel{\textstyle<}{\sim}\;$}}


\def\de{\partial}
\def\Tr{ \hbox{\rm Tr}}
\def\const{\hbox {\rm const.}}  
\def\o{\over}
\def\im{\hbox{\rm Im}}
\def\re{\hbox{\rm Re}}
\def\bra{\langle}\def\ket{\rangle}
\def\Arg{\hbox {\rm Arg}}
\def\Re{\hbox {\rm Re}}
\def\Im{\hbox {\rm Im}}
\def\diag{\hbox{\rm diag}}


\def\QATOPD#1#2#3#4{{#3 \atopwithdelims#1#2 #4}}
\def\stackunder#1#2{\mathrel{\mathop{#2}\limits_{#1}}}
\def\stackreb#1#2{\mathrel{\mathop{#2}\limits_{#1}}}
\def\Tr{{\rm Tr}}
\def\res{{\rm res}}
\def\Bf#1{\mbox{\boldmath $#1$}}
\def\balpha{{\Bf\alpha}}
\def\bbeta{{\Bf\beta}}
\def\bgamma{{\Bf\gamma}}
\def\bnu{{\Bf\nu}}
\def\bmu{{\Bf\mu}}
\def\bphi{{\Bf\phi}}
\def\bPhi{{\Bf\Phi}}
\def\bomega{{\Bf\omega}}
\def\blambda{{\Bf\lambda}}
\def\brho{{\Bf\rho}}
\def\bsigma{{\bfit\sigma}}
\def\bxi{{\Bf\xi}}
\def\bbeta{{\Bf\eta}}
\def\d{\partial}
\def\der#1#2{\frac{\d{#1}}{\d{#2}}}
\def\Im{{\rm Im}}
\def\Re{{\rm Re}}
\def\rank{{\rm rank}}
\def\diag{{\rm diag}}
\def\2{{1\over 2}}
\def\ntwo{${\mathcal N}=2\;$}
\def\nfour{${\mathcal N}=4\;$}
\def\none{${\mathcal N}=1\;$}
\def\ntwot{${\mathcal N}=(2,2)\;$}
\def\ntwoo{${\mathcal N}=(0,2)\;$}
\def\x{\stackrel{\otimes}{,}}

\def\ba{\beq\new\begin{array}{c}}
\def\ea{\end{array}\eeq}
\def\be{\ba}
\def\ee{\ea}
\def\stackreb#1#2{\mathrel{\mathop{#2}\limits_{#1}}}

\def\Tr{{\rm Tr}}
\newcommand{\cpn}{CP$(N-1)\;$}
\newcommand{\wcpn}{wCP$_{N,\tilde{N}}(N_f-1)\;$}
\newcommand{\wcpd}{wCP$_{\tilde{N},N}(N_f-1)\;$}
\newcommand{\vp}{\varphi}
\newcommand{\pt}{\partial}
\newcommand{\ve}{\varepsilon}
\renewcommand{\theequation}{\thesection.\arabic{equation}}

\setcounter{footnote}0

\vfill

\begin{titlepage}

\begin{flushright}
FTPI-MINN-13/10, UMN-TH-3142/13\\
\end{flushright}

\vspace{1mm}

\begin{center}
{  \Large \bf  
\boldmath{Detailing \none Seiberg's Duality through \\[1mm]the Seiberg-Witten Solution of \ntwo
 
}}

\vspace{5mm}

 {\large \bf    M.~Shifman$^{\,a}$ and \bf A.~Yung$^{\,\,a,b}$}
\end {center}

\begin{center}

$^a${\it  William I. Fine Theoretical Physics Institute,
University of Minnesota,
Minneapolis, MN 55455, USA}\\
$^{b}${\it Petersburg Nuclear Physics Institute, Gatchina, St. Petersburg
188300, Russia
}
\end{center}

\vspace{1mm}

\begin{center}
{\large\bf Abstract}
\end{center}

Starting from the Seiberg-Witten solution
of  \ntwo SQCD with the U$(N)$ gauge group and $N_f$ quark flavors  
we construct the so-called $\mu$-dual \none theory in the $r$ vacua in the regime 
analogous to that existing 
to the left of the left edge 
of the Seiberg conformal window, where $r$ is 
the number of condensed quarks. The strong-weak coupling duality is shown to exist 
in the so-called zero vacua which can be found  at $r< N_f-N$. We show that the $\mu$-dual theory matches 
the Seiberg dual in the zero vacua.

\vspace{2cm}

\end{titlepage}

 \newpage



\section {Introduction }
\label{intro}
\setcounter{equation}{0}

Seiberg's duality in its original formulation \cite{Sdual,IS} relates \none supersymmet\-ric QCD (SQCD) with the
SU$(N)$ gauge group and $N_f$ quark flavors  to a dual  theory
with the SU$(\tN)$ gauge group, the same number of dual quarks, plus 
a neutral meson field $M$. Here 
\beq
\tN\equiv N_f-N\,.
\label{nnn}
\eeq
These two theories forming the Seiberg pair are distinctly different in the ultraviolet (UV) domain, but describe exactly
the same dynamics in the infrared (IR) domain.
Later Seiberg's duality was generalized to other gauge groups and extended to other matter contents.
Although Seiberg's duality was a conjecture it passed numerous tests both on the field  and string theory sides, and is viewed as firmly established.
 
A breakthrough in understanding the strong coupling gauge dynamics was achieved with the Seiberg-Witten solution 
\cite{SW1,SW2}
  of \ntwo  SQCD. Combining the above two constructions together
 could shed light on the physical nature of Seiberg's dual quarks and provide us with an additional understanding of low-energy physics in \none SQCD, in particular, physics of confinement and screening in the regime where  
the dual theory is weakly coupled.

A crucial step in this direction was made in \cite{APS}. In this paper SU$(N)$ \ntwo theory deformed 
by the mass term $\mu\Tr\,{\mathcal A}^2$ for the adjoint matter was considered. At small $\mu$ this theory was described by
the Seiberg-Witten solution \cite{SW1,SW2}, while at large $\mu$ it obviously flows to \none
SQCD. It was shown that the SU$(\tN)$ gauge group present at low energies at the root of a baryonic branch survives the large $\mu$ limit. This explains the emergence of the SU$(\tN)$ gauge group in the Seiberg's dual theory.
The presence of a large number of distinct vacua in the IR, with different physical features, was not discussed in \cite{APS}.
And understandably so,  since the analysis of \cite{APS}
was carried out with massless quarks in which case certain vacua coalesce, and  Higgs branches develop from common roots.

Much later it was noted  (in the framework of the U$(N)$ gauge theories) that Seiberg's dual
theory and the theory at the baryonic root are associated with different vacua \cite{SYN1dual}.
To identify distinct vacua we introduced mass terms $m_A$, $A=1,...,N_f$
to the quark fields. It is known that the $\mu$-deformed \ntwo SQCD with generic quark masses has
the  so-called  $r$ vacua (they are isolated) in which $r$ quark flavors condense,\footnote{Note that 
$r=N$ is the maximum possible number of condensed quarks. The $r$ counting is carried our at large
$m_A$, see below.} $r\le N$. 
 The Seiberg \none duality  was in fact formulated for monopole vacua with $r=0$ in the limit 
 $\mu\to \infty$.
In the $r\ne 0$ vacua the condensates of $r$ quark flavors are determined by the value of the effective parameters
$\xi_A \sim \mu m_A$, hence, they are runaway vacua in the limit $\mu\to \infty$
corresponding to \none\!\!.
The root of the baryonic branch in the U$(N)$ version of the theory corresponds to the $r=N$ isolated vacuum.  In the limit $\mu\to \infty$ this vacuum becomes a
runaway vacuum too.

The number of quark flavors $N_f$ to be considered below is subject to the constraint
\beq
N+1< N_f < \frac32\, N\,.
\label{nconstr}
\eeq
This domain lies to the left of the left edge of the Seiberg conformal window. In this domain 
the original ``electric" theory in the Seiberg pair is asymptotically free and strongly coupled in the IR, while 
its dual ``magnetic" partner is infrared free and weakly coupled in the IR. This pattern will be preserved in our consideration. 

In this paper we mostly consider $r$ vacua with ``small" $r$,
\beq
r<\frac{N_f}{2},
\label{smallr}
\eeq
see \cite{SYrvacua,Shifman:2011ka} and Sec.~\ref{larr} for the discussion of $r>N_f/2$ vacua.
Our strategy is as follows: we start from the original U$(N)$ theory in the \none large-$\mu$ limit, which is in fact the UV limit of the theory. Then we decrease $\mu$ approaching the \ntwo limit.
At this stage the mass parameters $m_A$ are kept large.
Then we use the Seiberg-Witten solution to analytically continue to the domain of small $m_A$.  The theory obtained in this way still has \ntwo supersymmetry. Then we increase $\mu $ 
to decouple the adjoint scalar superfield and return to   \none\!. In doing so we keep $\mu$ large but finite
in order to keep track of all $r$ vacua.
In this limit we find an IR-free model,
the dual partner to our original \none theory. At every stage of this road full theoretical control is maintained, including the IR domain.
The Seiberg-Witten solution is combined with the powerful tools worked out by Cachazo, Seiberg, and Witten \cite{Cachazo2}, and by
Dijkgraaf and Vafa \cite{DijVafa}. This allows us to identify, from the analysis of the dual partners,  the relevant vacua and their dynamics. 
We are only interested in such dual partners that are at  weak coupling in the IR, thus maintaining the same pattern as 
the one inherent to the Seiberg duality in the domain to the left from the conformal window. We will see that the
original U($N$)
theory can flow in the IR to the  dual IR free theory, with the gauge group U($\tN$) (i.e. exactly the same as in \cite{Sdual,IS}), possessing  special $r$ vacua, to be referred to as the {\em zero vacua}. We discuss the corresponding dynamics, as well as the nature of Seiberg's dual quarks. 

To briefly explain the emergence and relevance of the zero vacua in the problem at hand we note that
   our starting point is \ntwo SQCD with a small $\mu\Tr\,{\mathcal A}^2$ term. In $r$ vacuum with  $r<N_f/2$ 
 at low energies, after developing condensates of $r$ quarks,  this theory reduces to the IR free 
U$(r)\times$U(1)$^{N-r}$ gauge theory with $r$ light quarks and $(N-r-1)$ Abelian monopoles.\footnote{The U$(r)$
gauge factor implies that all mass terms $m_A$ are almost equal.} Using the results of \cite{SW1,SW2,Cachazo2}
one can detect a large number of various $r$ vacua in the above low-energy theory.
Among these vacua we identify a special set of the
zero vacua, namely those,  in which the  gaugino condensate tends to zero in the small $m_A$ limit.
In all other $r$ vacua   (to be referred to as $\Lambda$ vacua) it stays finite. 
In fact, the zero  vacua exist only at
\beq 
  r<\tN\,.
\label{zve}
\eeq
  
\begin{figure}[h]
\epsfxsize=6cm
\centerline{\epsfbox{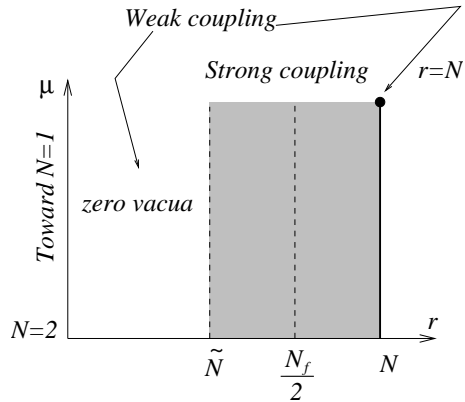}}
\caption{\small Uplifting \ntwo theory to \none. The zero vacua for which weakly coupled $\mu$-dual theories exists
can be found in the unshaded domain. The $r=N$ theory in the upper-right corner is exceptional. For
$r=N$  weakly coupled dual theory exists, while all other theories in the shaded domain have strongly-coupled duals.
}
\label{uplift}
\end{figure}

The above theory  can be ``uplifted" (by increasing $\mu$) to \none\!\!. 
This uplift leads to the original U$(N)$ theory in UV (see Fig. \ref{uplift}).
At the same time,  at small $m_A$ the uplift from the zero vacua leads us to an
\none  $\mu$-dual theory  weakly coupled in the IR and strongly coupled in the UV,  with the enhanced 
U($\tN$) gauge group and  $N_f$ flavors of quarks. The  $r$
 quark flavors condensed in the vacuum trigger confinement of monopoles charged with respect to the Cartan generators of the  SU$(r)$ group. Thus the dual theory is in the mixed Coulomb/Higgs phase. 
 The U$(\tN)$ gauge group of the $\mu$-dual theory is the same as  Seiberg's dual gauge group.
 We explicitly show  that the $\mu$-dual theory matches the generalized Seiberg dual in the zero 
 vacua.\footnote{The generalization of the Seiberg duality for all $r$ vacua in 
 $\mu$ deformed U$(N)$ SQCD (with finite $\mu$) 
was worked out  in \cite{CKM}, see also \cite{GivKut}.} This match reveals the nature of Seiberg's dual quarks. They are just ordinary quarks of the original theory.

What happens to the $\Lambda$ vacua, which exist both in the interval $N> r \geq \tN$ (populated exclusively  by such vacua)
and in the interval (\ref{zve})? These vacua do not have IR weak coupling  descriptions at large $\mu$.
 Unfortunately, this was overlooked in \cite{SYrvacua,SYvsS}, where we claimed a discrepancy between the
so-called $r$-dual theory and the generalized Seiberg dual  at large 
$\mu$.\footnote{A loophole was the assumption of weak coupling in the regime, which 
is a continuation of the Argyres-Douglas (AD) points \cite{AD} to large $\mu$, 
while in fact the regime considered was at strong coupling.}
Here we correct this claim. The only exception is the $r=N$ case, where a weakly 
coupled dual with the Seiberg U$(N_f-N)$ group does exist 
\cite{Shifman:2011ka}.

To conclude the introductory section
we note that previously we discussed \cite{SYdual,Shifman:2011ka} the $r>N_f/2$ vacua 
in the {\em \ntwo limit} in some detail.
In the  \ntwo limit (small $\mu$) the strong-coupling domain of
the original theory in the $r$ vacua (with $r>N_f/2$) can be described in terms of 
a weakly coupled $r$-dual theory. 
The gauge group in this theory is
U$(\nu)\times$U(1)$^{N-\nu}$, $\nu=N_f-r$.
Moreover, the $r$-dual theory has  $N_f$ flavors of quark-like dyons. Condensation of these dyons
leads to the confinement of monopoles. Quarks and gauge bosons of the original theory are in the 
``instead-of-confinement'' phase \cite{SYrvacua,SYvsS,SYdual,SYN1dual}. However, as was 
already mentioned above, 
this weak-coupling $r$-dual description present at small $\mu$ becomes strongly coupled  once we
increase $\mu$,  see Fig.~\ref{uplift}.

The paper is organized as follows. In Sec.~\ref{bulk} we review $\mu$-deformed \ntwo supersymmetric QCD
and its vacuum structure in the limit of small $\mu$. In Sec.~\ref{zerovacua} we identify zero and
$\Lambda$-vacua using Cachazo-Seiberg-Witten exact solution for chiral rings \cite{Cachazo2}.
In Sec.~\ref{muduality} we describe $\mu$-duality which relates the U$(r)\times$U(1)$^{N-r}$ 
hybrid quark-monopole low energy
theory present in zero vacua at small $\mu$ to $\mu$-dual quark theory with U($\tN$) gauge group
emerging at large $\mu$. In Sec.~\ref{seibergdual} we discuss the  generalization of Seiberg's duality to 
$r$ vacua of the theory at large but finite $\mu$ and show the match of $\mu$-dual theory
with Seiberg's dual. Finally, in Sec.~\ref{phases} we briefly describe $r$ duality in $r>N_f/2$ vacua and 
summarize various phases of \none QCD present in the $r$ vacua at strong coupling. In Sec.~8 we present our
conclusions.

\section {Preliminaries}
\label{bulk}
\setcounter{equation}{0}

\subsection{ \ntwo SQCD (small $\mu$)}
\label{bulkp}

Our basic ``microscopic" (or UV) theory is described in detail in our previous publications (e.g. \cite{SYmon,SYfstr}
and review papers (e.g. \cite{SYrev}), where the reader can find all relevant notation. 
The gauge symmetry  is 
U($N$)=SU$(N)\times$U(1), with the $\mu\Tr\,{\mathcal A}^2$ deformation term. We
have $N_f$  quark hypermultiplets generally speaking endowed with the mass terms $m_A$. 
The number of flavors is subject to the constraint (\ref{nconstr})
ensuring  asymptotic  freedom of the microscopic theory as well as IR freedom of the dual theory.

 The  superpotential of the undeformed  \ntwo theory has the form
 \beq
{\mathcal W}_{{\mathcal N}=2} = \sqrt{2}\,\sum_{A=1}^{N_f}
\left( \frac{1}{ 2}\,\tilde q_A {\mathcal A}
q^A +  \tilde q_A {\mathcal A}^a\,T^a  q^A + m_A\,\tilde q_A q^A\right)\,,
\label{superpot}
\eeq
where ${\mathcal A}$ and ${\mathcal A}^a$ are  chiral superfields, the ${\mathcal N}=2$
superpartners of the U(1) and SU($N$) gauge bosons. The deformation term
\beq
{\mathcal W}_{{\rm def}}=
  \mu\,{\rm Tr}\,\Phi^2, \qquad \Phi\equiv\frac12\, {\mathcal A} + T^a\, {\mathcal A}^a
\label{msuperpotbr}
\eeq
does not break \ntwo supersymmetry in the small-$\mu$ limit,  see \cite{HSZ,VY,SYmon}
(while at large $\mu$ this theory obviously flows to \none). For small $\mu$,
i.e. $\mu\ll \Lambda_{{\mathcal N}=2}$,   and if all quark masses are equal this term reduces to the 
Fayet-Iliopoulos $F$ term  which can be rotated  \cite{HSZ,VY,SYmon,SYfstr} into the $D$ term \cite{FI}.

\subsection{Vacua}
\label{vacu}

We define the $r$ vacuum as a vacuum with $r$ flavors of (s)quarks condensed
It is assumed that the $r$ counting is performed 
 at large quark masses. As we will see in Sec. 3,
 effectively the value of $r$   depends on the quark masses \cite{SYvsS}. It is obvious that
the maximal value of  $r$ is $N$. If $r=N$ the gauge group is fully Higgsed  \cite{SYrev,SYdual}.

For generic $m_A$   
the number of the isolated $r$ vacua with $r<N$  is \cite{CKM}
\beq
{\cal N}_{r<N}=\sum_{r=0}^{N-1} \,(N-r)\,C_{N_f}^{r}= \sum_{r=0}^{N-1}\, (N-r)\,\frac{N_f!}{r!(N_f-r)!}\,.
\label{nurvac}
\eeq

Consider a
 particular vacuum in which the first $r$ quarks develop nonvanishing vacuum expectation values (VEVs). 
Quasiclassically, at large masses, the adjoint scalar VEVs   are  
\beq
\left\langle \Phi\right\rangle \approx - \frac1{\sqrt{2}}\,
{\rm diag}\left[m_1,...,m_r,0, ..., 0 
\right],
\label{avevr}
\eeq
The last $(N-r)$ entries vanish at the   classical level.
In quantum theory these entries acquire values of the order of $\Lambda_{{\mathcal N}=2}$, generally speaking.
In the classically unbroken U$(N-r)$ pure gauge sector the gauge symmetry gets broken through the Seiberg--Witten mechanism \cite{SW1}:
first down to U(1)$^{N-r}$ and then almost completely by condensation of $(N-r-1)$ monopoles. A single
 U(1) gauge factor survives, though,
because monopoles are charged only with respect to 
the Cartan generators of the SU$(N-r)$ group.

The presence of this unbroken U(1) factor
in all $r<N$ vacua makes them different from the $r=N$ vacuum: in the latter there are
no long-range forces. 

In this paper we  focus on the $r$ vacua with $r< N_f/2$.
Then the low-energy theory in the given $r$ vacuum (following from the microscopic theory under consideration) has the
\beq
{\rm U}(r)\times {\rm U}(1)^{N-r}\,,
\label{legaugegroup}
\eeq
gauge group, assuming that the quark masses are almost equal. Moreover, 
$N_f$  quarks are charged
under the U$(r)$ factor, while  $(N-r-1)$ monopoles are charged under the U(1) factors. Note that the
quarks and monopoles are charged with respect to orthogonal subgroups of U$(N)$
and therefore are mutually local (i.e. can be described by a local Lagrangian). 
The low-energy theory
is infrared-free and it is at  weak coupling as long as VEVs of quarks and monopoles are small. 

\subsection{Large values of \boldmath{$m_A$}}

The quark VEVs in the large-mass limit can be read off from the superpotentials (\ref{superpot}) and (\ref{msuperpotbr}) using (\ref{avevr}). They are given by
\beqn
\langle q^{kA}\rangle &=& \langle\bar{\tilde{q}}^{kA}\rangle=\frac1{\sqrt{2}}\,
\left(
\begin{array}{cccccc}
\sqrt{\xi_1} & \ldots & 0 & 0 & \ldots & 0\\
\ldots & \ldots & \ldots  & \ldots & \ldots & \ldots\\
0 & \ldots & \sqrt{\xi_r} & 0 & \ldots & 0\\
\end{array}
\right),
\nonumber\\[4mm]
k&=&1,..., r\,,\qquad A=1,...,N_f\, ,
\label{qvevr}
\eeqn
where the $r$ parameters $\xi$ are given quasiclassically by
\beq
\xi_P \approx 2\;\mu m_P,
\qquad P=1,..., r\,.
\label{xiclass}
\eeq
These parameters can be made small in the large $m_A$ limit if $\mu$ is sufficiently small.

In quantum theory all parameters $\xi_P$  are determined by the roots of the Seiberg-Witten curve
\cite{SYfstr,SYN1dual,SYrvacua,SYhybrid} which in the case at hand 
 takes the form \cite{APS}
\beq
y^2= \prod_{P=1}^{N} (x-\phi_P)^2 -
4\left(\frac{\Lambda_{{\mathcal N}=2}}{\sqrt{2}}\right)^{2N-N_f}\, \,\,\prod_{A=1}^{N_f} \left(x+\frac{m_A}{\sqrt{2}}\right).
\label{curve}
\eeq
Here $\phi_P$ are gauge invariant parameters on the Coulomb branch. Instead of (\ref{avevr}) one can write
\beq
\Phi \approx 
{\rm diag}\left[\phi_1,...,\phi_N\right],
\eeq
where
\beq
\phi_P \approx -\frac{m_P}{\sqrt{2}},\quad P=1, ... ,\, r\,; \qquad 
\phi_P \sim \Lambda_{{\mathcal N}=2},\quad P=r+1, ... ,\, N\,.
\label{classphi}
\eeq

To identify the $r$ vacuum in terms of the curve (\ref{curve}) it is necessary to find
such values of $\phi_P$ which ensure the Seiberg-Witten curve to have $N-1$ double roots, while 
$r$ parameters $\phi_P$ are {\em approximately} determined by the quark masses, see (\ref{classphi}).
Note that $(N-1)$ double roots are associated with $r$ condensed quarks and $(N-r-1)$ condensed monopoles,
 altogether $N-1$ condensed states.

From this we deduce that the Seiberg--Witten curve factorizes \cite{CaInVa},
\beq
y^2
=\prod_{P=1}^{r} (x-e_P)^2\,\prod_{K=r+1}^{N-1} (x-e_K)^2\,(x-e_N^{+})(x-e_N^{-})\,.
\label{rcurve}
\eeq
The first $r$ double roots are associated with  the
mass parameters in the large mass limit, $\sqrt{2}e_P\approx  -m_P$, $P=1, ... , r$. The subsequent $(N-r-1)$ double roots are associated with light monopoles are much smaller, and determined by $\Lambda_{{\mathcal N}=2}$.
The last two  roots  are also much smaller. 
 For the single-trace deformation superpotential (\ref{msuperpotbr}) 
 their sum vanishes \cite{CaInVa},
\beq
e_N^{+} + e_N^{-}=0\,.
\label{DijVafa}
\eeq
The root $e_N^{+}$ determines the value of the gaugino condensate \cite{Cachazo2},
\beq
e_N^2=\frac{2S}{\mu}, \qquad S=\frac1{32\pi^2}\langle {\rm Tr}\,W_{\alpha}W^{\alpha} \rangle,
\label{eN}
\eeq
where the superfield $W_{\alpha}$ includes the gauge field strength tensor.

In terms of roots of the Seiberg-Witten curve the quark VEVs  are given by 
the formula \cite{SYrvacua,SYhybrid}
\beq
\xi_P=-2\sqrt{2}\,\mu\,\sqrt{(e_P-e_N^{+})(e_P-e_N^{-})}
\label{xi}
\eeq
for $P=1,...,r$.
At small $\xi_P$ this theory is at weak coupling (IR free below $\Lambda_{{\mathcal N}=2}$) and 
supports non-Abelian magnetic strings
 \cite{HT1,ABEKY,SYmon,HT2}. At  $\mu\ll \Lambda_{{\mathcal N}=2}$ these strings are BPS-saturated and their tensions 
 are determined by the  $\xi$ parameters, namely  \cite{SYrev,SYfstr}
 \beq
T_P=2\pi|\xi_P|.
\label{mten}
\eeq
Magnetic strings formed as a consequence of the  quark condensation implement confinement of monopoles. The monopoles
of the SU$(r)$ sector manifest themselves as two-string junctions \cite{SYmon,HT2,SYtorkink}.

Recently we demonstrated \cite{SYhybrid}
that the monopole VEVs in either the monopole ($r=0$) or the hybrid $r$ vacua 
are determined by the same
formula (\ref{xi}) with the substitutions of the quark double roots by the monopole double roots, 
so that the subscript $P$ in (\ref{xi}) can run
over the monopole double roots too,
\beq
  \langle M_{P(P+1)}\rangle = \langle\bar{\tilde{M}}_{P(P+1)}\rangle = \sqrt{\frac{\xi_P}{2}},  
 \label{mvev}
 \eeq
where $\xi_P$ are determined by Eq. (\ref{xi}) and  $P=(r+1),..., (N-1)$. Here $M_{PP'}$ denotes the  monopole with the charge given by  the root $\alpha_{PP'}=w_P-w_{P'}$ of 
the 
SU$(N)$ algebra with weights $w_P$ ($P<P'$).

Equation (\ref{xi}) is thus very general and determines
VEVs of  condensed states independently of their nature \cite{SYhybrid}.
The monopole VEVs determine the tensions of the Abelian electric strings, 
\beq
T_P=2\pi|\xi_P|, \qquad P=(r+1),..., (N-1)\,.
\label{eten}
\eeq
In much in the same way as the magnetic non-Abelian strings  in the $r$ vacua,
the electric strings are BPS-saturated to the leading order in $\mu$ \cite{HSZ,VY}. The electric strings
confine quarks, while the magnetic strings confine monopoles.

\subsection{Small quark mass limit}
\label{smalmq}

Now we turn to the opposite limit of small $m_A$  which will be relevant to our discussion below.
 As we reduce the quark masses  quantum numbers of the light states change due to 
 monodromies \cite{SW1,SW2,BF}. In particular,
quarks pick up root-like color-magnetic charges, in addition to their weight-like color-electric charges.
If  $r< N_f/2 $  there is no crossover, the low-energy theory essentially remains the same as at large $m_A$, namely,
infrared-free U$(r)\times$U(1)$^{N-r}$ gauge theory with $N_f$ quarks (or, more exactly, what becomes of quarks)
and $(N-r-1)$ singlet monopoles \cite{MY2}. It is at weak coupling provided the $\xi_P$ parameters  are small.

The quarks from the U$(r)$ sector and the monopoles form the orthogonal U(1)$^{N-r}$ sector still develop VEVs 
determined by Eq. (\ref{xi}). 
Physics of screening and confinement also remains intact at small $m_A$. Say, if a given monopole state 
(charged with respect to the SU$(r)$ Cartan generators) is confined through quark condensation at large $m_A$ the the same applies to this state under the evolution into the domain of  small $m_A$,  although the quark color charges change \cite{MY2}. If the quarks from the U$(r)$ sector are screened in the $r$ vacuum at large $m_A$
 they (or what becomes of them) will still be screened in the same vacuum at small $m_A$.
 Monodromies  just relabel the  states, they do not change  physics. 

\section { \boldmath{$\Lambda$} vacua versus  zero vacua}
\label{zerovacua}
\setcounter{equation}{0}

\subsection{Consequences from the exact formulas}

We will rely on exact results  for the chiral condensates 
obtained by Cachazo, Seiberg and Witten \cite{Cachazo2} in $\mu$-deformed \ntwo QCD with 
the U$(N)$
gauge group. In this section there is no need to assume $\mu$ small.

All chiral condensates are encoded in the following functions \cite{Cachazo2}:
\beqn
&&
T(x)=\left\langle {\rm Tr}\,\frac1{x-\Phi}\right\rangle,
\nonumber\\[2mm]
&&
R(x)=\frac1{32\pi^2}\,\left\langle {\rm Tr}\,\frac{W_{\alpha}W^{\alpha}}{x-\Phi}\right\rangle,
\nonumber\\[2mm]
&&
M(x)_A^B=\left\langle \tilde{q}_A\,\frac1{x-\Phi}\,q^B\right\rangle.
\label{rings}
\eeqn
For the quadratic single-trace deformation (\ref{msuperpotbr}) (the so-called ``one-cut'' model) the function
$R(x)$ has the form
\beq
R(x)=\frac12\left(W_{\rm def}^{'}(x) -\sqrt{W_{\rm def}^{'}(x)+f(x)}\right)
=\mu\,\left(x-\sqrt{x^2-e_N^2}\right),
\label{R(x)}
\eeq
where the unpaired root of the Seiberg--Witten curve $e_N = e_N^{+}$ (see (\ref{rcurve})) is related to the gaugino condensate, see (\ref{eN}).

From the solution for the function $M_A^B(x)$ in \cite{Cachazo2} one can obtain the values of 
the quark VEVs
in terms of the gaugino condensate $S$.
In the $r$ vacuum, when the function $M_A^B(x)$ has $r$ poles on the first sheet,  
\beqn
&&
M_A =  \frac{\mu}{2}\, \left(m_A+\sqrt{m_A^2-\frac{4S}{\mu}}\right), \qquad A=1,..., r\,,;
\nonumber\\[2mm]
&&
M_A =  \frac{\mu}{2}\, \left(m_A-\sqrt{m_A^2-\frac{4S}{\mu}}\right), \quad A=(r+1),..., N_f, 
\label{qVEVcsw}
\eeqn
where  
\beq
M_A^B=\left\langle \tilde{q}_A q^B\right\rangle\,,
\label{M}
\eeq
 and we assume that the solution can be brought to the diagonal form 
 \beq
 M_A^B=\delta_A^B\,M_A\,.
 \eeq
In the large quark mass limit, when $\frac S\mu\ll m_A$ , we have 
 $r$ ``large" values of $M_A$,
 $$M_A\approx \mu\, m_A\,\,\, \mbox{ for}\,\,\, A=1,..., r\,,$$ and $N_f- r$ ``small" values.
 This pattern matches  our definition of the $r$ vacuum.
 
 Note, that the quantum quark VEVs (\ref{M}) in  the microscopic  U$(N)$ theory  
 are close to those obtained from
 the low-energy theory (see  (\ref{qvevr}))  only in the limit of  large $m_A$, although both are given by
 exact formulas. At $m_A\sim \Lambda_{{\mathcal N}=2}$ the difference is not small.
 In particular, the low-energy quark condensates (\ref{qvevr}) vanish at the Argyres-Douglas
 points \cite{AD} (where a double root $e_P$ coincides with one of the unpaired roots $e_N^{\pm}$), see (\ref{xi}),
 while the values $M_A$ remain finite. This was first noted in \cite{GVY}.

Now, to find the gaugino condensate $S$ we use the glueball superpotential calculated in \cite{Cachazo2} from
a matrix model \cite{DijVafa}. For the quadratic deformation (\ref{msuperpotbr}) it was studied in 
\cite{Ookouchi}, see also \cite{SYvsS}.
Minimization of this superpotential gives the following equation for $S$: 
\beq
S^N= \mu^N\,\Lambda_{{\mathcal N}=2}^{N-\tN}\,\left(\frac{m}{2}-\frac12\sqrt{m^2-\frac{4S}{\mu}}\right)^r
\,\left(\frac{m}{2}+\frac12\sqrt{m^2-\frac{4S}{\mu}}\right)^{N_f-r},
\label{Seqn}
\eeq
where for simplicity we assume quark mass equality.
Using  (\ref{qVEVcsw}) we can rewrite the equation above as an equation for the quark condensate
$M_A$ \cite{SYvsS},
\beq
\frac1{\mu}\,M_A = m -\frac1{\mu^{\frac{N}{\tN}}\,\Lambda_{{\mathcal N}=2}^{\frac{N-\tN}{\tN}}}\,
\frac{\left({\rm det}\,M \right)^{\frac1{\tN}}}{M_A},
\label{qeqCSW}
\eeq
where $\tN$ is defined in (\ref{nnn}). 
Equation (\ref{qeqCSW}) obviously can be obtained from the following superpotential:
\beq
{\mathcal W}_{\rm ADS} = -\frac{1}{2\mu}\,{\rm Tr}\,M^2 + m_A\, {\rm Tr}\,M 
+(N-N_f)\,\frac{\left({\rm det}\,M\right)^{\frac1{N_f-N}}}{\Lambda^{\frac{3N-N_f}{N_f-N}}}
\,.
\label{ads}
\eeq
The first two terms in (\ref{ads})  can be obtained by integrating out the adjoint field ${\mathcal A}$ in the tree-level
superpotential of the theory  (\ref{superpot}) and (\ref{msuperpotbr}) in the large $\mu$
limit.
The last  term  -- obviously of the quantum nature -- is nothing other than the continuation of the
Afleck-Dine-Seiberg (ADS) superpotential  \cite{ADS} to $N_f>N$. This superpotential can be also derived from Seiberg's dual theory generalized to
$r$ vacua, see \cite{SYvsS} and Sec.~\ref{seibergdual} below.

Thus,  
the Cachazo--Seiberg--Witten
exact solution \cite{Cachazo2} produces the same equations for 
$M$'s as the  continuation of 
the ADS superpotential to $N_f>N$ in Eq.~(\ref{ads}). 
The fact of coincidence was previously established in the
 SU$(N)$ case in \cite{Pietro}.

The superpotential (\ref{ads}) is exact and we can use it in any domain of the parameter space. In particular,
for large masses ($m_A\gg \Lambda_{{\mathcal N}=2}$) the solution of Eq.~(\ref{qeqCSW}) in the 
$r$ vacuum
is
\beqn
&&
M_A \approx  \mu\, m, \qquad A=1,..., r\,;
\nonumber\\[3mm]
&&
M_A \approx  \mu\, \Lambda_{{\mathcal N}=2}^{\frac{N-\tN}{N-r}}\;
m^{\frac{\tN-r}{N-r}}\;e^{\frac{2\pi k}{N-r}\,i}, \qquad A=(r+1),..., N_f, 
\nonumber\\[4mm]
&&
 k=1,... , (N-r)\,.
\label{qvevS}
\eeqn
 As was anticipated, we  have $r$ large classical VEVs and $(N_f-r)$ small 
``quantum'' VEVs.

The linear dependence of $M$ on $\mu$ is exact and is fixed by the U(1) symmetries \cite{GVY}
after all condensates are expressed in terms of $\Lambda_{{\mathcal N}=2}$. The presence of 
$(N-r)$ distinct solutions ensures  the total number of the $r<N$ vacua to  coincide with
(\ref{nurvac}) obtained at small $\mu$.

\subsection{Chiral condensates at small quark masses}
\label{ccc}

Let us study the behavior of gaugino  and quark condensates at small $m_A$. Most of the solutions
of equations (\ref{Seqn}), (\ref{qVEVcsw}) behave as $S\sim \mu \,\Lambda_{{\mathcal N}=2}^2$ and 
$M_A\sim \mu \,\Lambda_{{\mathcal N}=2}$. The  $r$ vacua with this behavior are referred to as the $\Lambda$
vacua. However, there is a special set of vacua in which the gaugino and quark condensates
tend to zero in the small quark mass limit. Namely, Eq.~(\ref{qeqCSW}) has solutions \cite{GivKut,SYvsS}
\beqn
&&
M_A  \approx  \mu\, m, \qquad A=1,..., p\,;
\nonumber\\[2mm]
&&
M_A  \approx  \mu\, 
\frac{m^{\frac{p-\tN}{p-N}}}{\Lambda_{{\mathcal N}=2}^{\frac{N-\tN}{p-N}}}\;e^{\frac{2\pi k}{p-N}\,i}\,, \qquad A=(p+1),...,N_f,\; 
\nonumber\\[4mm]
&&
 k=1,... , (p-N)\,,
\label{smallqvev}
\eeqn
where $p$ is an integer. In other words,  $p$ eigenvalues of $M$ are proportional to $\mu\,m$, 
while other eigenvalues are much smaller at $m_A\approx m \ll \Lambda_{{\mathcal N}=2}$.
These solutions exist if $p>N$. We refer to the vacua with this behavior as the zero vacua.

At large $m_A$ we start  from an $r$ vacuum,  with $r$ quarks (classically) condensed, hence
$r\le N$. On the other hand, the integer $p$ is defined as 
the number of ``plus'' signs in Eq.~(\ref{qVEVcsw}) for $M_A$, or the number of poles of $M_A^B(x)$ on the
first sheet \cite{Cachazo2}.
Then $(N_f-p)$ is the number of ``minus'' signs. In fact, $p$ depends on the value of $m_A$. At large $m_A$ we have
\beq p(\infty)=r\,.\eeq
 As we reduce $m_A$ certain poles can and do pass through the cut from the first sheet to the second or
vice versa \cite{Cachazo2}. When it happens $p(m_A)$ reduces by one unit or increases by one unit. 

In Eq.~(\ref{smallqvev}) 
$p$ is $p(m_A)$ in the small mass limit, i.e. 
\beq p\equiv p(0)\,.\eeq 
Clearly, $p$ can differ from $r$. 
The condition $r \le N$  applies only for $r=p(\infty)$ rather than for $p=p(0)$, instead
 $p>N$. In fact, $(p-r)$
is the net number of poles which pass through the cut from the second sheet to the first one as we reduce 
the quark masses from infinity to zero. 

The relation between $r$ and $p$ was found in \cite{SYvsS}. We look for a solution of  
(\ref{qeqCSW}) which has the pattern (\ref{qvevS}) at large $m$ and  (\ref{smallqvev}) at small $m$.
Translating this into the behavior of $S$ given by Eq.~(\ref{Seqn}) we arrive at \cite{SYvsS}
\beq
p=N_f-r.
\label{pr}
\eeq
Then constraint $p>N$ implies in turn that
\beq
r< \tN\,.
\label{rtN}
\eeq
This is the domain of existence of the zero vacua.

\begin{figure}
\epsfxsize=8cm
\centerline{\epsfbox{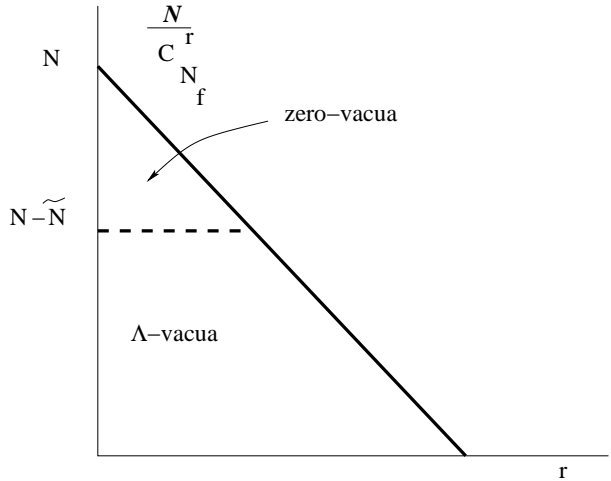}}
\caption{\small Multiplicity of zero-vacua and total multiplicity of $r<N$ vacua as a function of $r$.}
\label{fig:zerovac}
\end{figure}

Equation (\ref{smallqvev}) for $M_A$ in the zero vacua ensures the smallness of the  gaugino condensate
 at small $m_A$,
\beq
S\approx  \mu\, 
\frac{m^{\frac{N_f-2r}{\tN -r}}}{\Lambda_{{\mathcal N}=2}^{\frac{N-\tN}{\tN-r}}}\;e^{\frac{2\pi k}{\tN-r}\,i}\,, 
\qquad 
 k=1,... , (\tN-r)\,,
\label{0vacgaugino}
\eeq
where we express $p$ in terms of $r$ using (\ref{pr}).
The multiplicity of these solutions is $\tN-r$. In other words,  for a given $r$  the total number of the zero vacua is
\beq
{\cal N}_{0-{\rm vac}}=\sum_{r=0}^{\tN-1} \,(\tN-r)\,C_{N_f}^{r}=
\sum_{r=0}^{\tN-1}\, (\tN-r)\,\frac{N_f!}{r!(N_f-r)!}\,.
\label{nu0vac}
\eeq
This number is smaller then the total number of the $r<N$ vacua (\ref{nurvac}). The multiplicity
of the zero vacua as a function of $r$ is depicted in Fig.~\ref{fig:zerovac}.

We will show below that choosing any of the zero vacua
we can pass from the weak coupling low-energy description of Sec.~\ref{bulk}  at small $\mu$ (i.e. \ntwo) 
to a $\mu$-dual
 \none theory which appears to be weakly coupled in the IR. At the same time, the zero vacua 
were shown 
\cite{GivKut,SYvsS} to be precisely the vacua which are
classically seen in the generalized Seiberg dual  theory \cite{Sdual,IS}.   
Section \ref{seibergdual} elucidates that these are two sides of the same coin.

\section {Towards {\boldmath\none}\! by increasing \boldmath{$\mu$}:\\ \boldmath{$\mu$} Duality }
\label{muduality}
\setcounter{equation}{0}

\subsection{Preliminaries}
\label{prel}

 If the quark mass differences are small $(m_A-m_B) \ll m_{A,B} \sim m \ll \Lambda_{{\mathcal N}=2}$ 
 then $r$ parameters $\phi_P$ and 
 the quark double roots $e_P$  (in $r<N_f/2$ vacua) are exactly (rather than quasiclassically) determined 
by the
quark masses \cite{APS,SYdual,SYrvacua}, 
\beq
\sqrt{2}\,\phi_P = -m_P, \qquad \sqrt{2}\, e_P = -m_P, \qquad P=1,...,r
\label{rroots}
\eeq
(up to a small corrections of order of $(m_A-m_B)^2/\Lambda_{{\mathcal N}=2}$). The 
Seiberg-Witten curve factorizes as follows
\beq
y^2= \left(x+\frac{m}{\sqrt{2}}\right)^{2r}\left\{\prod_{P=r+1}^{N} (x-\phi_P)^2 -
4\left(\frac{\Lambda_{{\mathcal N}=2}}{\sqrt{2}}\right)^{2N-N_f}\, \,\, 
\left(x+\frac{m}{\sqrt{2}}\right)^{N_f-2r} \right\}.
\label{factcurve}
\eeq
This leads to the occurrence of the non-Abelian SU($r$) gauge group in the low-energy theory in the limit of (almost)
equal quark masses \cite{APS}.
As a result, at small $\mu$  physics is described by weakly coupled IR free 
low-energy theory discussed in Sec.~\ref{bulk}.
It has the U$(r)\times {\rm U}(1)^{N-r}$ gauge group  with $r$ light  quarks and $(N-r-1)$
Abelian monopoles. 

Our task is to increase $\mu$ and find a weakly coupled
low-energy description of the theory at hand at large $\mu$ (i.e. \none). However, this program runs onto an obstacle.
At large $\mu$ the $\xi$ parameters  (\ref{xi}) generically become large forcing the infrared-free low-energy theory hit the strong coupling domain.

Previously we believed  \cite{SYrvacua} that the problem could be overcome by approaching the Argyres-Douglas  points
\cite{AD} where $r$  double roots  come close to one of the 
unpaired roots $e_N^{\pm}$ and $r$  parameters $\xi$ remain small. 
It was overlooked, however,  that the low-energy theory in this limit enters the AD strongly coupled regime,
while our task was to find a weakly coupled dual.\footnote{In \cite{SYrvacua,SYvsS} it was argued  that 
the low-energy theory stays
at weak coupling near the AD points at  $\mu\,\, \slash\!\!\!\!\!\!\!\to 0$ because the monopoles which become light are, in fact, confined and, therefore, do not contribute to the $\beta$ function. The loophole in this argument is that at energies above the
scale $\sqrt{\xi}$ the effect of confinement is negligible, and the light monopoles do cancel the logarithmic running of the coupling constant produced by the light quarks. We will discuss  this issue in more detail elsewhere.}

One exception where this problem does not appear is the $r=N$ vacuum. In the 
$r=N$ vacuum the gaugino condensate
vanishes, and $\tN=N_f-N$ parameters $\xi$ are determined by the quarks masses \cite{SYN1dual}, $$\xi_P=-2\sqrt{2}\,\mu m_P\,,\quad
P=1,...,\tN\,.$$ This allows us 
to keep the $\xi_P$ parameters  small at large $\mu$ by making the
quark masses sufficiently small, guaranteeing  a weak coupling regime in the   dual theory
which in this case has the U$(\tN)$ gauge group \cite{SYN1dual}, in perfect agreement with Seiberg's duality.

Now we want to demonstrate that  the zero vacua provide us with additional exceptions. The gaugino condensate is very small in the limit
of small masses, see (\ref{0vacgaugino}). Therefore we do not need to approach the AD points to keep $r$ parameters $\xi$ small. What we need is  to  make the quark masses  small as we increase $\mu$.

\subsection{The zero vacua from the Seiberg-Witten curve}
\label{zvsw}

We begin with   identifying the  zero vacua in terms of the Seiberg-Witten curve (\ref{factcurve}).
The gaugino condensate is related to values of the unpaired roots of the curve (see (\ref{eN})),

\beq
e_N^2 \approx  2\mu\, 
\frac{m^{\frac{N_f-2r}{\tN -r}}}{\Lambda_{{\mathcal N}=2}^{\frac{N-\tN}{\tN-r}}}\;e^{\frac{2\pi k}{\tN-r}\,i}\,, 
\qquad 
 k=1,... , (\tN-r)\,,
\label{0vaceN}
\eeq
in the small mass limit. All other roots of the curve (\ref{factcurve}) are doubled.
In the zero vacua  $r$ parameters $\phi_P$ and the double roots $e_P$ are given by the quark masses, $P=1,..., r$
(see (\ref{rroots})), while 
$(N-\tN)$ parameters $\phi_P$ and the double roots $e_P$ are of order of $\Lambda_{{\mathcal N}=2}$. 
The remaining $(\tN-r)$ parameters $\phi$ and $(\tN-r-1)$ double roots are very small,
of the order of $e_N^{\pm}$, see (\ref{0vaceN}).

To find $\phi$'s and those double roots which are of the order of $\Lambda_{{\mathcal N}=2}$ (and for this purpose only) we consider $x\sim \Lambda_{{\mathcal N}=2}$ in (\ref{factcurve}) and  neglect all parameters which are of the order of $m$ or smaller
(remember that  $m\ll\Lambda_{{\mathcal N}=2}$).
Then, Eq. (\ref{factcurve}) implies
\beq
y^2= x^{2\tN}\left\{\prod_{P=\tN+1}^{N} (x-\phi_P)^2 -
4\left(\frac{\Lambda_{{\mathcal N}=2}}{\sqrt{2}}\right)^{N-\tN}\, \,\, 
x^{N-\tN} \right\}.
\label{Lcurve}
\eeq
We look for a solution with all $(N-\tN)$ $\phi$'s being of the order of $\Lambda_{{\mathcal N}=2}$.
 Of course, there are solutions with  smaller $\phi$'s given by the quark masses,
but these solutions correspond to $r'$-vacua with larger $r'$, i.e.  $r'>\tN$.

The solution takes the form
\beq
\rule{5mm} {0mm}
\sqrt{2}\,\phi_P  =  -\Lambda_{{\mathcal N}=2}\,e^{\frac{2\pi i}{N-\tN}(P-\tN-1)}, \quad P=(\tN+1),..., N, \quad\mbox{odd $(N-\tN)$}, 
\label{phiodd}
\eeq
 and 
\beq
\sqrt{2}\,\phi_P  =  -\Lambda_{{\mathcal N}=2}\,e^{\frac{2\pi i}{N-\tN}(P-\tN-\frac12)}, \quad P=(\tN+1),..., N,\quad
\mbox{even $(N-\tN)$}\,.
\label{phieven}
\eeq
The corresponding double roots are\,\footnote{Note a shift in the numbering
of the ``large" $\phi$'s and the double roots: the double root $e_P$ corresponds to $\phi_{P+1}$. This is because
we use the notation $e_N^{\pm}$  for unpaired roots, which are small.}
\beq
\sqrt{2}\,e_P  =  \Lambda_{{\mathcal N}=2}\,e^{\frac{2\pi i}{N-\tN}(P-\tN)}, \qquad P=\tN,...,(N-1).
\label{Lroots}
\eeq

To find the remaining $(\tN-r)$  $\phi$'s and the roots that are much smaller than $m$ we assume in (\ref{factcurve}) 
$x\ll m$. The Seiberg-Witten curve then  takes the form
\beq
y^2= \left(\frac{m}{\sqrt{2}}\right)^{2r}\,
\left(\frac{\Lambda_{{\mathcal N}=2}}{\sqrt{2}}\right)^{2(N-\tN)}
\left\{\prod_{P=r+1}^{\tN} (x-\phi_P)^2 \,-
4\frac{ \left(\frac{m}{\sqrt{2}}\right)^{N_f-2r}}{\left(\frac{\Lambda_{{\mathcal N}=2}}{\sqrt{2}}\right)^{N-\tN}} \right\},
\label{smallxcurve}
\eeq
where we use the fact that $(N-\tN)$ $\phi$'s are given by (\ref{phiodd}) or (\ref{phieven}).
The curve in the curly brackets   is the curve for  pure Yang-Mills theory with the U$(\tN-r)$ gauge group.
It has a very small scale $\Lambda_0$ defined as
\beq
\Lambda_0^{2(\tN-r)} = \frac{m^{N_f-2r}}{\Lambda_{{\mathcal N}=2}^{N-\tN}}\, , \qquad \Lambda_0 \ll m\,.
\label{Lambda0}
\eeq
The relevant
parameters $\phi$ as well as the roots in pure Yang-Mills theory were obtained in \cite{DS},
\beq
\phi_P=2 \cos{\frac{\pi(P-r-\frac12)}{\tN-r}}\,\frac{\Lambda_0}{\sqrt{2}},
\qquad P=(r+1),...,\tN\,,
\label{smallphi}
\eeq
and
\beq
e_P=2 \cos{\frac{\pi(P-r)}{\tN-r}}\,\frac{\Lambda_0}{\sqrt{2}},
\qquad P=(r+1),...,(\tN-1)\,.
\label{smalldoubleroots}
\eeq
The unpaired roots are
\beq
e_N^{\pm}=\pm 2 \frac{\Lambda_0}{\sqrt{2}}\,.
\label{smalleN}
\eeq

Comparing the above expression for the unpaired roots  found from the Seiberg-Witten curve with
the result (\ref{0vaceN}) obtained using the Cachazo-Seiberg-Witten exact solution \cite{Cachazo2},
applied to the  zero vacua, we observe the exact match.

Next,  can  use (\ref{smalldoubleroots}) and (\ref{smalleN}) to find ``small" VEVs for $(\tN-r-1)$ monopoles. 
They are given by  (\ref{mvev}), where for $P=(r+1),...,(\tN-1)$,
\beq
\xi_P=-2\sqrt{2}\,\mu\,\sqrt{(e_P-e_N^{+})(e_P-e_N^{-})}=-4 i\,\mu\Lambda_0\,
\sin{\frac{\pi(P-r)}{\tN-r}},
\label{sine}
\eeq
see (\ref{xi}). This is the famous sine formula for the monopole VEVs and the Abelian electric string
tensions \cite{DS}. We reproduce it via our general expression (\ref{xi}) which we can use 
in particular, for pure Yang-Mills 
theory, see  \cite{SYhybrid}. Note that $r$ quark VEVs (\ref{qvevr}) are determined by the quark masses,
\beq
\xi_P = 2\mu m_P, \qquad P=1,....,r\,,
\label{rxis}
\eeq
since $e_N^{\pm}$ are very small at small $m$, see (\ref{xi}) and (\ref{rroots}).
Other $(N-\tN)$ monopoles have ``large" VEVs determined by
$\Lambda_{{\mathcal N}=2}$, see (\ref{Lroots}). These monopoles decouple from   low-energy physics.

\subsection{\boldmath{$\mu$} Dual theory in the zero vacua}

The above analysis implies that the low-energy theories in the zero vacua at small $\mu$
and $m$ have the U$(r)\times$U(1)$^{\tN-r}$ gauge group with $r$ flavors of light quarks charged under
the U($r$) subgroup  and $(\tN-r-1)$ light monopoles charged under the $(\tN-r-1)$ U(1) factors. 
One U(1) remains unbroken. The remaining
U(1)$^{N-\tN}$ gauge sector becomes heavy and decouples, along with $(N-\tN)$ heavy monopoles.
 
The  scale $\Lambda_0$ of the U(1)$^{\tN-r}$ sector is very small, as it is clearly seen from (\ref{Lambda0}).
Therefore, when we increase
$\mu$ forcing  VEVs (\ref{sine})  of $(\tN-r-1)$ light monopoles to hit the scale $\Lambda_0$,
 the U(1)$^{\tN-r}$
monopole sector enters the strong coupling regime, and we cannot use this monopole  theory to describe
  low-energy physics.

Nevertheless, at larger $\mu$ we can construct a dual 
low-energy description. Equations (\ref{avevr}), (\ref{phiodd}), (\ref{phieven}) and (\ref{smallphi}) show
that the adjoint field in  the zero vacuum has the form
\beq
\left\langle \Phi\right\rangle \approx - \frac1{\sqrt{2}}\,
{\rm diag}\left[m_1,...,m_r,0,...,0, c_1\Lambda_{{\mathcal N}=2},...,c_{N-\tN}\Lambda_{{\mathcal N}=2} 
\right],
\label{avevzero}
\eeq
where we have $(\tN-r)$ almost vanishing eigenvalues, while $(N-\tN)$ ``large" entries (i.e. of the order of 
$\Lambda_{{\mathcal N}=2}$)  are associated with  the decoupled U(1)$^{N-\tN}$ heavy sector.
The form of the adjoint field in (\ref{avevzero}) signals the restoration of the U($\tN$) gauge group
(if $m\ll \Lambda_{{\mathcal N}=2}$).

The Seiberg-Witten curve  takes the form
\beq
y^2= \left(x+\frac{m}{\sqrt{2}}\right)^{2r}\,
\left(\frac{\Lambda_{{\mathcal N}=2}}{\sqrt{2}}\right)^{2(N-\tN)}
\left\{\prod_{P=r+1}^{\tN} (x-\phi_P)^2 \,-
4\frac{ \left(x+\frac{m}{\sqrt{2}}\right)^{N_f-2r}}{\left(\frac{\Lambda_{{\mathcal N}=2}}{\sqrt{2}}\right)^{N-\tN}} \right\}.
\label{mudualcurve}
\eeq
We focus on the low-energy region, $x\ll\Lambda_{{\mathcal N}=2}$.
This is the curve of the IR free U($\tN$) gauge theory with $N_f$ flavors.
Thus, the $\mu$-dual low-energy theory has the U($\tN$) gauge group
and $N_f$ quark flavors. The superpotential of the theory is
\beqn
{\mathcal W}_{\mu-{\rm dual}} & = & \sqrt{2}\,\sum_{A=1}^{N_f}
\left( \frac{1}{ 2}\,\tilde q_A {\mathcal A}
q^A +  \tilde q_A {\mathcal A}^n\,T^n  q^A + m_A\,\tilde q_A q^A\right)
\nonumber\\[2mm]
&+& \mu\,u_2,
\label{mudualsuperpot}
\eeqn
where
\beq
u_2={\rm Tr}\,\left(\frac12\, {\mathcal A} + T^n\, {\mathcal A}^n\right)^2,
\label{u2}
\eeq
while the fundamental and adjoint color indices are now truncated to $l=1,...,\tN$ and $n=1,...,(\tN^2-1)$.

The VEVs of the adjoint field are
\beq
\left\langle \frac12\, {\mathcal A} + T^n\, {\mathcal A}^n\right\rangle \approx - \frac1{\sqrt{2}}\,
{\rm diag}\left[m_1,...,m_r,0,...,0
\right],
\label{avevzerotrnk}
\eeq
where $(\tN-r)$ eigenvalues  are quasiclassically zero. The matrix of the quark VEVs has the form
\beqn
\langle q^{lA}\rangle &=& \langle\bar{\tilde{q}}^{lA}\rangle=\frac1{\sqrt{2}}\,
\left(
\begin{array}{cccccc}
\sqrt{\xi_1} & \ldots & 0 & 0 & \ldots & 0\\
0 & \ldots & \ldots  & \ldots & \ldots & 0\\
0 & \ldots & \sqrt{\xi_r} & 0 & \ldots & 0\\
0 & \ldots & 0 & 0 & \ldots & 0\\
0 & \ldots & \ldots & \ldots & \ldots & 0\\
0 & \ldots & 0 & 0 & \ldots & 0
\end{array}
\right),
\nonumber\\[4mm]
l&=&1,..., \tN\,,\qquad A=1,...,N_f\, ,
\label{qvevzero}
\eeqn
where the first  $r$ parameters $\xi$ are given  by (\ref{rxis}), while all other $(\tN-r)$ $\xi$'s are (quasiclassically) zero.
Only $r$ quark flavors  develop VEVs. The U($\tN-r$) gauge sector remains unbroken. The U($\tN$) theory is IR free   and is weakly coupled at energies  above $\Lambda_0$, see below.

Let us stress that the reason why the low-energy superpotential (\ref{mudualsuperpot}) is consistent with
the adjoint and quark VEVs given above is a peculiar property  of the zero vacuum namely, the extreme smallness
of $(\tN-r-1)$ parameters $\xi$ which are of the order of $\Lambda_0\ll m$, see (\ref{sine}).
Generically, if only $r$ quarks of $\tN$ (the maximal possible value allowed by the rank of the gauge group)
condense, the $F$ terms proportional to
$$
\mu\,\frac{\pt u_2}{\pt \Phi} \sim \xi
$$
are generated. To cancel these terms, additional $(\tN-r-1)$ monopoles develop VEVs. 

The reason why this does not happen in the zero vacua  at $\mu\gg \Lambda_0$ is 
the fact that the corresponding 
parameters $\xi$ are (almost) zero, see (\ref{sine}).

At energies  above $m$ the U$(\tN)$ $\mu$-dual theory at hand is IR free and weakly coupled.
However, at energies below $m$ the gauge group gets broken to U$(r)\times$U$(\tN-r)$ by adjoint VEVs
(\ref{avevzerotrnk}).
The U$(r)$ sector with $N_f$ light quarks is IR free and weakly coupled. However, the 
U$(\tN-r)$ sector becomes a pure Yang-Mills theory since the quarks charged with respect to 
U$(\tN-r)$ gauge group acquire masses of the order of $m$ and decouple.\footnote{This is because of extreme smallness of the corresponding parameters $\phi$, see (\ref{smallphi}).} Thus, the U$(\tN-r)$ sector
is asymptotically free and runs into strong coupling in the infrared. This happens at the scale of
U$(\tN-r)$ Yang-Mills theory, which coincides with $\Lambda_0$, see (\ref{Lambda0}).

Thus, we must admit that the  $\mu$-dual theory at hand is not exactly a  
weakly coupled low-energy description all the way down. It is weakly coupled only at energies  above the very small scale provided by $\Lambda_0\ll m$.

 At energies below $\Lambda_0$ we have
a weak coupling description in terms of the  U$(r)\times$U(1)$^{\tN-r}$ gauge theory with light quarks and monopoles. As we increase $\mu$ and go to higher energy scales our system undergoes  
a crossover transition, and the quark-monopole description breaks down. 

At energies well above $\Lambda_0$
we use weakly coupled $\mu$-dual description in terms of the U($\tN$) gauge theory for $N_f$ light quarks.
This is quite natural because the monopoles are Abelian objects and hardly can play a role at large $\mu$
where adjoint fields decouple and no Abelianization of the theory is expected.

\subsection{Superpotential}
\label{superp}

Masses of quarks and gauge bosons in the $\mu$-dual theory are determined by the scales $m$ and $\sqrt{\xi}$
in (\ref{rxis}). Therefore, once we increase $\mu$ above $m$ the adjoint fields decouple from  low-energy physics
making the theory at hand \none. It is easy to integrate out adjoint fields in the 
superpotential (\ref{mudualsuperpot}).
We expand $u_2$ in $a$ and $a^n$ and keep only quadratic terms (higher order terms are suppressed 
by powers of $m/\Lambda_{{\mathcal N}=2}$). The coefficients of this expansion are determined by using 
the adjoint and quark VEVs (\ref{avevzerotrnk}) and (\ref{qvevzero}). In this way we get
\beq
{\mathcal W}_{\mu-{\rm dual}} = -\frac1{2\mu}\,(\tilde{q}_Aq^B)(\tilde{q}_Bq^A) + m_A \tilde{q}_Aq^A\,,
\label{sup}
\eeq
for further details see  \cite{SYN1dual} where a similar calculation is carried out  in the  $r=N$ vacuum.

To summarize, at  large $\mu$ the original U$(N)$ gauge theory flows to \none SQCD.
At  $\mu\gg m$  low-energy physics in the  zero vacua can be described by
\none supersymmetric SQCD with the
U($\tN$) gauge group and $N_f$ quark flavors   with the superpotential (\ref{sup}).
Note, that in contrast to \ntwo SQCD,  where in each vacuum we have its own description with a distinct gauge group, in the case at hand we have one and the same superpotential for all zero vacua. The vacua differ by the number $r$
of condensed quarks. In order to keep this IR free theory at weak coupling we assume that the $\xi$ 
parameters in (\ref{rxis}) are small compared to the scale of this \none $\mu$-dual theory, determined by
\beq
\tilde{\Lambda}_{{\mathcal N}=1}^{N_f-3\tN}=\frac{\Lambda_{{\mathcal N}=2}^{N_f-2\tN}}{\mu^{\tN}}.
\label{N1Lambda}
\eeq
Namely, we assume
\beq
\xi \sim \mu m \ll \tilde{\Lambda}_{{\mathcal N}=1}^2.
\label{mudualregion}
\eeq

Condensation of quarks leads to the formation of non-Abelian strings in the U$(r)$ sector.
These strings  confine monopoles, for a review on non-Abelian strings and monopole confinement see \cite{SYrev}. 
The U$(\tN-r)$ sector remains unbroken. Thus, our theory is in the mixed Higgs/Coulomb phase.
Quarks of the U$(r)$ sector are screened, while monopoles are confined. 

We stress that quarks are not confined at large $\mu$, contrary to the naive duality arguments.

To conclude, let us address the question: how large should $\mu$  be to ensure the decoupling of the adjoint matter?
From (\ref{N1Lambda}) we see that in order to make contact with \ntwo theory we cannot take $\mu$ too
large; in fact, it cannot exceed $\Lambda_{{\mathcal N}=2}$. This upper bound might seem too 
restrictive from the point of view of the original 
microscopic \none U$(N)$ SQCD. Indeed one might think that in order 
to decouple the adjoint matter one should take $\mu$ much larger than the 
scale of this theory. However,  above we saw that the low-energy states in the $\mu$-dual 
theory have masses determined by $m$ or $\sqrt{\xi}$ which are way below the 
above-mentioned scale. Therefore, in order to decouple 
the adjoint matter from the low-energy sector 
it is sufficient to keep $\mu$ in the window $m \ll \mu \lsim \Lambda_{{\mathcal N}=2}$.

\section {Connection to Seiberg's duality }
\label{seibergdual}
\setcounter{equation}{0}

As was mentioned in Sec.~\ref{intro},
originally Seiberg's duality \cite{Sdual,IS} was formulated for \none SQCD   corresponding to the limit $\mu\to \infty$
and referred to  $r=0$. 
A  generalization of Seiberg's duality for $r$ vacua of  $\mu$-deformed \ntwo SQCD at large but
finite $\mu$ was considered in \cite{CKM,GivKut}. In our case of the U$(N)$ SQCD  the Seiberg's dual
has the U$(\tN)$ gauge group,  $N_f$ flavors of Seiberg's dual quarks\,\footnote{To be  denoted as  $h^{lA}$ ($l=1,...,\tN$ and $A=1,...,N_f$).} and neutral mesonic field $M_A^B$ defined in (\ref{M}). The Seiberg superpotential is
\beq
{\cal W}_{S}= -\frac{1}{2\mu}\,{\rm Tr}\,(M^2) + m_A\, M_A^A +\frac1{\kappa}\,\tilde{h}_{Al}h^{lB}\,M_B^A\,,
\label{Ssup}
\eeq
where first two terms are obtained by integrating out the adjoint fields at  the tree level in 
(\ref{superpot}) and (\ref{msuperpotbr}). Here $\kappa$ is a parameter of dimension of mass needed to formulate Seiberg's duality \cite{Sdual,IS}.

From  definition (\ref{M}) it is clear that 
the number of the eigenvalues of the matrix $\tilde{q} q = M$ which scales as
$ \mu m$ at large $  m$ is $r$ in the $r$ vacuum. 
What is the vacuum structure \cite{CKM,GivKut,SYvsS} of the Seiberg dual theory (\ref{Ssup}) for the $r<N$ vacua?  

If we integrate out Seiberg's dual quarks $h^{lA}$ we end up \cite{IS,CKM,GivKut,SYvsS} with the Afleck-Dine-Seiberg superpotential (\ref{ads}). It correctly reproduces the total number of the $r<N$ vacua (\ref{nurvac}) and gives 
the correct values of the $M$ condensates since Eq. (\ref{qeqCSW})
coincides with the one obtained from the Cachazo-Seiberg-Witten exact solution \cite{Cachazo2}, see \cite{SYvsS} and Sec.~\ref{zerovacua}. However, the ADS superpotential (\ref{ads}) is not a superpotential of a gauge theory (gauge degrees of freedom are already integrated out). In fact this is a superpotential of the 
Veneziano-Yankielowicz type \cite{Ven} and, as such, cannot be used to  describe  the spectrum of low-energy excitations, confinement or screening \cite{SYvsS}. It is useful only for the vacuum condensates.

To describe   low-energy physics we need 
a weakly coupled description in terms of a gauge theory.  We could try to use the Seiberg dual theory (\ref{Ssup})
{\em per se}. In \cite{GivKut} it was   noted that 
not  all $r<N$ vacua  
can be seen at the classical level in the superpotential  (\ref{Ssup}).
Later  it was found \cite{SYvsS} that only the zero vacua are  seen in (\ref{Ssup})  at the classical level,
while the $\Lambda$ vacua remain ``missing,'' or  quantum vacua, seen only in the ADS superpotential (\ref{ads}).
Let us briefly discuss this.

Extremizing superpotential (\ref{Ssup}) we find the classical vacua of 
the generalized Seiberg dual theory. Assuming that $\langle M_A^B\rangle =\delta_A^B\,M_A$
we arrive at
\beqn
&&
-\frac{1}{\mu}\,M_A + \kappa\,m_A +\frac1{\kappa}\,\tilde{h}_{Al}h^{lA}=0,
\nonumber\\
&&
M_A\,h^{lA}=\tilde{h}_{Al}\,M_A =0,
\label{veqs}
\eeqn
for all values of $A$.
The solution of (\ref{veqs}) is
\beqn
&&
  M_A = \mu\, m_A, \qquad (\tilde{h}h)_A=0, \qquad A=1,... , p\,,
\nonumber\\[2mm]
&&
(\tilde{h}h)_A = -\kappa\,m_A, \qquad M_A=0,  \qquad A=(p+1),... , N_f\,,
\label{GKclvac}
\eeqn
where $p$ should obey the constraint $p>N$, since the rank of the matrix $(\tilde{h}h)$ cannot exceed $\tN$.

This solution can describe low-energy physics if the infrared-free Seiberg  dual theory is at
weak coupling. To ensure that this is the case we assume the small-$m$ limit. 
 In this limit $p$  does not coincide with $r$, the latter  parameter being 
defined at large masses. In fact $p=N_f-r$, see \cite{SYvsS} and (\ref{pr}). Now observe that 
$p$ eigenvalues of $M$ are given by $\mu m$, while others are classically zeros.
This dependence matches the $m$ dependence of $M$ in the zero vacua at small $m$, see (\ref{smallqvev}).
Moreover, the number of classical vacua (\ref{GKclvac}) is 
\beq
{\cal N}_{0-{\rm vac}}=\sum_{r=0}^{\tN-1} \,(\tN-r)\,C_{N_f}^{r}=
\sum_{r=0}^{\tN-1}\, (\tN-r)\,\frac{N_f!}{r!(N_f-r)!}\,.
\label{nuclSvac}
\eeq
This is the number of choices one can pick up $r=N_f-p$ dual quarks $h$ which develop VEVs times the Witten index  in the classically unbroken by  $h$ condensation gauge group, namely
SU($\tN-r$). This number coincides with the zero vacua number, see (\ref{nu0vac}) and Fig.~\ref{fig:zerovac}. 

This leads us to the conclusion that vacua (\ref{GKclvac}) classically seen in the Seiberg dual theory are 
in fact the zero vacua \cite{SYvsS}. The $\Lambda$ vacua are not seen classically. Our interpretation of this 
phenomenon is as follows (cf. \cite{SYvsS}).  In the zero vacua U$(\tN)$ is the true low-energy gauge group and dual quarks $h$ are the correct low-energy degrees of freedom. Since the Seiberg dual theory is infrared-free 
it is weakly coupled in the small-$m$ limit, provided the classical vacua  exist, i.e. in the zero vacua.
Instead, in the  $\Lambda$ vacua, the dual quarks $h$ are {\em not} the low-energy degrees 
of freedom. 

This explains why the $\Lambda$ vacua are not seen quasiclassically. In fact,
Seiberg's dual U$(\tN)$ theory (\ref{Ssup}) is strongly coupled in the $\Lambda$ vacua. Nevertheless, integrating out
dual quarks leads to the correct ADS superpotential (\ref{ads}), which can be used only to determine 
chiral condensates from the chiral rings, \`a la Veneziano-Yankielowicz.

In much the same way as in the Seiberg duality, our $\mu$-dual theory in the zero vacua  
 also has the U$(\tN)$ gauge group (Sec.~\ref{muduality}).
Both dual theories give  weakly coupled low-energy descriptions in the small-$m$ limit.
Do  these two descriptions match?

The answer is positive. To see that this is the case, let us    identify the quarks of 
the $\mu$-dual theory with the Seiberg dual quarks. The change of variables
\beq
q^{lA}=\sqrt{-\frac{\mu}{\kappa}}\, h^{lA}\,, \quad N_A^B\equiv -\frac1{\mu}\, M_A^B\,, \quad l=1,...,\tN, \quad A=1,...,N_f
\label{change}
\eeq
brings the superpotential (\ref{Ssup}) to the form 
\beq
{\cal W}_{S}= -\frac{\mu}{2}\,{\rm Tr}\,(N^2) -\mu m_A\, N_A^A + \tilde{q}_{Al}q^{lB}\,N_B^A.
\label{Ssupch}
\eeq
The kinetic terms are not known in the Seiberg dual theory, and, hence,
 normalization of the $h$ fields is  unknown too, which leaves us the freedom to change the variables 
 as in (\ref{change}). We see that the $\kappa$ parameter completely
disappears from the theory and  is replaced by the physical parameter $\mu$. Equation (\ref{Ssupch})
shows that the mesonic field $N_A^B$ is heavy at large $\mu$ (i.e. $\mu \gg m$) and can be integrated out.
The result is 
\beq
{\mathcal W}_{S} = \frac1{2\mu}\,(\tilde{q}_Aq^B)(\tilde{q}_Bq^A) - m_A \tilde{q}_Aq^A.
\label{Ssupfin}
\eeq

This superpotential coincides with the superpotential (\ref{sup}) up to a sign. This shows the equivalence
of the Seiberg dual and $\mu$-dual low-energy theories in the zero vacua. The identification (\ref{change}) reveals the physical nature of Seiberg's dual quarks.
They are not monopoles as naive duality suggests. Instead,
 they are quarks of the original theory. Remember, $r$ quarks condense in the $r$ vacuum, see (\ref{qvevzero}). This  leads to confinement of monopoles charged with respect to the Cartan generators of SU$(r)$. Quarks of U($\tN-r$ ) sector do not condense, the dual theory is in the mixed Coulomb/Higgs phase.

\section {Phases of \none QCD in the small $\xi$ limit }
\label{phases}
\setcounter{equation}{0}

Before discussing the phases of \none SQCD we briefly review  the $r$ vacua with $r>N_f/2$ at small $\mu$.

\subsection{A few words about ``large"-\boldmath{$r$ ($r>N_f/2$}) vacua at small $\mu$}
 \label{larr}
 
  In the $r$ vacua with $r>N_f/2$ physics is quite different, see \cite{SYdual,SYrvacua,SYhybrid}.
  At large $\mu m$ ($\mu$ is assumed to be small so that the quark masses must be large) the 
  low energy-theory has the gauge group U$(r)\times$U(1)$^{N-r}$ 
  with $r$ condensed quarks and $(N-r-1)$ condensed monopoles. 
  The theory is at weak coupling because it has large condensates in the non-Abelian 
  asymptotically free SU$(r)$ quark sector and small condensates in the IR free monopole sector.
At low $\xi$ the theory goes through a crossover transition. At small $\xi$  physics can be described by
weakly coupled infrared-free $r$ dual theory with the U$(\nu)\times$U(1)$^{N-\nu}$ gauge group, $\nu = N_f-r$.
The $r$ dual theory has  $N_f$ flavors of quark-like dyons. The color charges of non-Abelian quark-like dyons 
are identical to those of quarks.\footnote{Because of monodromies quarks pick up root-like color-magnetic 
charges in addition to their weight-like color-electric charges at strong coupling.} However, they belong to a
different representation of the global color-flavor locked group.
Condensation of these dyons
leads to the confinement of monopoles.
Quarks from the U$(\nu)$ sector are in the ``instead-of-confinement'' phase: the Higgs-screened quarks decay into 
monopole-antimonopole pairs confined by non-Abelian strings. Singlet  quarks from  the U(1)$^{r-\nu}$ sector
and monopoles from the U(1)$^{N-r}$ sectors are Higgs-screened. Other monopoles, charged with 
respect to the Cartan generators of SU$(r)$, and quarks 
charged with respect to the orthogonal U(1)$^{N-r}$ are confined.

\subsection{Phases in  the $r$ vacua at large $\mu$}

\begin{figure}[h]
\epsfxsize=8cm
\centerline{\epsfbox{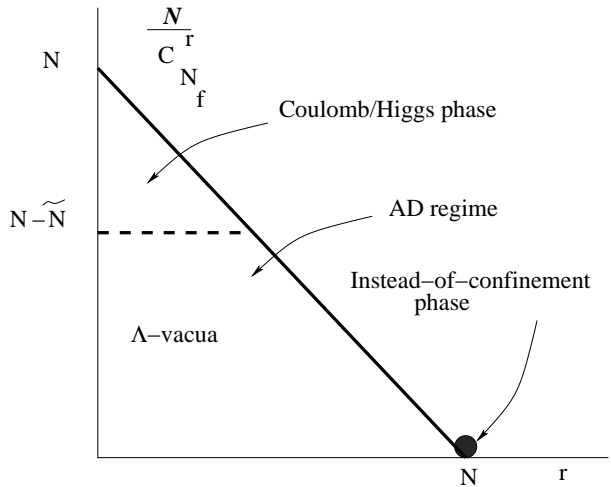}}
\caption{\small Phases of $r$ vacua in \none SQCD in the region (\ref{region}). Zero and 
$\Lambda$-vacua shown as in Fig.~\ref{fig:zerovac}. Black circle denotes $r=N$ vacua.
}
\label{fig:rphases}
\end{figure}

In this section we briefly summarize the overall  picture of  physical phases in different $r$ vacua 
 in the small $\xi$ and large $\mu$ limit. Namely, we impose
\beq
\mu\gg \sqrt{\xi}, \qquad 
\sqrt{\xi} \ll \tilde{\Lambda}_{{\mathcal N}=1}.
\label{region}
\eeq
Phases of the theory in the different $r$ vacua are shown in 
Fig.~\ref{fig:rphases}, which is the same as Fig.~\ref{fig:zerovac}, with various physical regimes indicated.

a) \underline{{\em Zero vacua}}: \hspace{2mm}  The $r$ parameters $\xi$ relevant to the low-energy $\mu$-dual theory 
are given by (\ref{rxis}).
This theory has the U$(\tN)$ gauge group with $N_f$ flavors of quarks.
It is infrared-free and weakly coupled in the region (\ref{region}) if we 
keep quark masses sufficiently small. The theory is 
in the mixed Coulomb/Higgs phase with $r$ quarks condensed, see (\ref{qvevzero}), while the U($\tN-r$ )
subgroup remains unbroken. Non-Abelian strings are formed in the U$(r)$ sector which entails confinement of monopoles charged with respect to the SU$(r)$ Cartan generators. The Seiberg and    $\mu$-dual descriptions are equivalent.

b) \underline{{\em $\Lambda$ vacua}}: \hspace{2mm}  As we increase $\mu$, we break the condition (\ref{region}),
generally speaking. 
The weak-coupling description is unknown so far, and it is unclear whether or not it exists. 
However,  we can tune the common quark mass $m$ and approach the Argyres-Douglas (AD) point
\cite{AD} where  $r$  double roots for $r<N_f/2$ vacuum and $\nu=N_f-r$ double roots  for $r> N_f/2$ vacuum  
come close to one of the  unpaired roots $e_N^{\pm}$. Then $r$ parameters $\xi$ for  
$r<N_f/2$ vacuum and $\nu=N_f-r$ parameters $\xi$  for $r> N_f/2$ vacuum 
can be made small to satisfy the bound in (\ref{region}), see (\ref{rroots}) and  \cite{SYrvacua}.
This limit is a continuation of the AD conformal 
{\em strongly coupled} regime to large $\mu$.

c) \underline{{\em The $r=N$ vacuum}}: \hspace{2mm}   
The $r=N$ vacuum presents a special case.
In this case the gaugino condensate vanishes
and $\tN$ parameters $\xi_P$ are proportional 
to $\mu m_P$. They can 
satisfy the bound (\ref{region}) provided the quark masses are sufficiently small.
The small-mass limit can be described by weakly coupled infrared-free 
$r$-dual theory \cite{SYN1dual,SYrvacua}. It has the U$(\tN)$ gauge group with $N_f$ flavors of quark-like
dyons. 
The quark-like dyons condense leading to the formation
of non-Abelian strings which confine monopoles. The 
quarks and gauge bosons of the original theory  are in 
the ``instead-of-confinement'' phase. Namely,
the Higgs-screened quarks and
gauge bosons decay into the monopole-antimonopole 
pairs on the curves of marginal stability (CMS) \cite{SYdual,SYtorkink}.
The monopole-antimonopole  pairs are in the confining regime. In other words, the original quarks and gauge bosons 
evolve at small  $\mu m$  into the monopole-antimonopole stringy mesons. (The latter are expected to form 
Regge trajectories, generally speaking). At $r=N$ the $r$-dual theory matches the Seiberg dual \cite{SYvsS}.
Conceptually this vacuum can be added to the zero vacua to form a class of vacua with U$(\tN)$ weak coupling
low-energy description. Moreover,  the number of condensed quark-like dyons for this vacuum is $\tN$ so 
it nicely adds to the set of zero vacua where the number of condensed quarks is $0\le r <\tN$.

\vspace{3mm}

To conclude this section, let us note that
there is no quark confinement   phase in \none SQCD in the domain (\ref{region}).
The Seiberg-Witten phase of monopole condensation and Abelian quark confinement present in the slightly
deformed \ntwo QCD at small $\mu$ \cite{SW1,SW2} does {\em not} survive in the 
large-$\mu$ domain where the adjoint fields 
decouple. This result resolves the long-standing problem of extrapolating the Seiberg-Witten scenario
of quark confinement to \none SQCD. The phase most close to what we observe in 
the real-world QCD is the ``instead-of-confinement'' phase present in the $r=N$ vacuum.

\section{Conclusions}

 We considered \ntwo SQCD with the U$(N)$ gauge group and $N_f$ quark flavors 
($N+1< N_f < \frac32\, N$) perturbed by 
a mass term $\mu{\mathcal A}^2$.  This theory has   $r$ vacua, i.e. those vacua in which
  $r$ flavors of quarks  condense, $r<N$ (this definition refers to large values of $m_A$).  In this paper
  we analyzed the $r$ vacua with $r<N_f/2$.
Low-energy theory in these vacua at small
$\mu$ is based on the U$(r)\times$U(1)$^{N-r}$ gauge group, with $r$ light quarks and $(N-r-1)$ Abelian monopoles.

Among these vacua we identify a subset that we call   {\em
zero vacua}. In the zero vacua the  gaugino condensate vanishes in the small quark-mass limit. We show that upon increasing $\mu$ these vacua go though a crossover into strong coupling.

At large $\mu$ the zero vacua can be described in terms of
weakly coupled infrared-free $\mu$-dual theory with the U($N_f-N $) gauge group and  $N_f$ flavors of quarks.
The $r$ quark flavors  condense triggering monopole confinement. We show that this $\mu$-dual theory matches the Seiberg  dual. This match reveals the nature of Seiberg's dual quarks which in this regime happen 
to be ordinary quarks of our microscopic theory  flowing to \none SQCD at large $\mu$.

The above conclusions are reached on the basis of the analysis of
the exact Seiberg-Witten solution
of  \ntwo SQCD. We focused on the 
  $\mu$-dual \none theories in the $r$ vacua in the regime analogous to that 
existing to the left of the left edge of the Seiberg conformal window. The strong-weak coupling duality is shown to exist 
in the zero vacua which can be found  at $r< N_f-N$.

\section*{Acknowledgments}
This work  is supported in part by DOE grant DE-FG02-94ER40823. 
The work of A.Y. was  supported 
by  FTPI, University of Minnesota, 
by RFBR Grant No. 13-02-00042a 
and by Russian State Grant for 
Scientific Schools RSGSS-657512010.2.

\end{document}